\documentclass[11pt]{article}

\newcommand{\be}{\begin{equation}}
\newcommand{\ee}{\end{equation}}
\newcommand{\bea}{\begin{eqnarray}}
\newcommand{\eea}{\end{eqnarray}}
\newcommand{\ba}{\begin{array}}
\newcommand{\ea}{\end{array}}

\begin{document}

\thispagestyle{empty} \vspace{40pt} \hfill{}

\vspace{128pt}

\begin{center}
\textbf{\Large ``So what will you do if string theory is wrong?'' }\\

\vspace{40pt}

Moataz H. Emam\footnote{Electronic address: memam@clarku.edu}

\vspace{12pt} \textit{Department of Physics}\\
\textit{Clark University}\\
\textit{Worcester, MA 01610}\\
\end{center}

\vspace{40pt}

\begin{abstract}

I briefly discuss the accomplishments of string theory that would
survive a complete falsification of the theory as a model of
nature and argue the possibility that such a survival may
necessarily mean that string theory would become its own
discipline, independently of both physics and mathematics.

\end{abstract}

\newpage


String theory occupies a special niche in the history of science.
It is the only theory of physics with no experimental backing that
has managed to not only survive, but also become ``the only game
in town'' (to quote Sheldon Glashow \cite{Greene:1999kj}). In
addition, the theory has gained much popularity with the general
public, spurred on by accessible online accounts and popular TV
programs. Judging by amateur web sites and personal discussions,
there seems to be a rising belief that it is a correct theory of
nature.

Of course, no one knows that for sure yet. This confusion has
extended even to the string physicists themselves. Although when
pressed they will tell you that string theory is still in the
hypothesis stage, many do act and talk as if it were confirmed
that it is a correct theory of the universe. This attitude has
triggered much criticism. One of the most vocal string theory
critics is Lee Smolin, although he has passed through a string
theory phase. In his book \cite{Smolin:2006pe} Smolin points out
that this ``faith'' in the string hypothesis has affected funding
and hiring policies in a negative way, effectively boosting the
theory's prominence disproportionately compared to other
approaches to quantum gravity \cite{Smolin:2000af}. Another
notable string theory critic is Peter Woit whose views can be
found in his book \cite{Woit:2006js}, articles
\cite{Woit:2002vr,Woit:2001jy}, and web site \cite{Woit:weblog}.
Although there are many counterarguments to be made, it seems that
string theory does receive more hype than it deserves if evaluated
solely on its applicability to nature and connection to
experiment.

In fact, string theory has so far failed to conform to the
definition of a scientific theory. In his classic work
\cite{Popper:1963} Karl Popper gives several criteria that a
scientific theory must satisfy. These may be summarized as ``the
criterion of the scientific status of a theory is its
falsifiability, or refutability, or testability''. A discussion
may be found in his cited original work as well as online sources
such as \cite{Popper1}. So far string theory has failed to meet
Popper's criterion. It might be argued that this situation is
temporary. Eventually technology will catch up with string theory
and allow us to test its assumptions directly or someone will find
a way to test the theory using current technology. This hope is
what keeps string theory on the list of scientific theories,
saving it from the fate of astrology and creationism. The failure
to satisfy Popper's definition is however a serious drawback that
string theory critics will, justly, continue to point out.

So why do people continue to work on string theory? There are
several reasons. We often hear that the theory is aesthetically
attractive and that it would be a shame if nature had not picked
such an elegant structure to use as the basis of the universe.
Furthermore, it is the only model that aspires to not just be a
theory of quantum gravity, but also a theory of everything;
unifying, in principle, all of known physics. The hope is that
eventually we will have a complete nonperturbative quantum theory
which leads to the standard model plus general relativity in the
low energy, dimensionally reduced limit. Not only that, we would
like this reduction to happen in a unique way. However, the
possible ways we can dimensionally reduce the ten-dimensional
string theory to four spacetime dimensions allows for many
possible outcomes, so large that they are collectively known as
the ``string theory landscape'' \cite{Banks:2003es,Bousso:2004fc}.
The often quoted estimate of the number of these product theories
is $10^{500}$! If the physics we observe is just one of $10^{500}$
possibilities, what of the remaining $\left(10^{500}-1\right)$
wrong ones? Why are they there? To date this is an open question.
An added complication is that the well-studied portions of the
landscape are far from being perfectly connected to each other.

The current situation of the theory may be likened to that of a
large beautiful Persian rug that is being woven thread by thread.
Usually we would start at some point in the rug and work our way
systematically through the elaborate designs such that at any
given time, we can see the completed portion of the rug all at
once. Unfortunately for the case of string theory, the unfinished
parts are not all in one place; they are scattered all over the
rug. It requires a considerable strain on the imagination to
visualize what the finished rug might eventually look like, or if
the completed pieces will ever smoothly and continuously meet.

People like myself who are interested in some small segment of the
string theory landscape that might not relate to the universe
naturally are asked: ``Why do you work on this theory? Shouldn't
you, as a physicist, be interested in what describes nature? Why
waste your time on something that you know \emph{a priori} to be
wrong?'' Another closely related question is ``What if someone
proves that subatomic particles cannot possibly be made of
strings? In that case not only is the particular theory you are
working on wrong, the whole edifice has collapsed! What will you
do then? Will you drop your research and switch to something else?
Or will you stubbornly continue to work on the (now incorrect)
string hypothesis? What will happen to all of your careers? And
why take the risk in the first place?'' These questions are
reasonable and may be rephrased as ``Are there any accomplishments
of string theory that would survive such a total collapse?'' It
turns out that there are.

The lack of experimental results to guide us through the vast
string landscape leaves string theorists with no choice but to
systematically explore all of it! These explorations, even within
theories that we already know are not related to nature, have
resulted in the discovery of deep and elegant mathematics.
Mathematicians today work in parallel with string theorists to
explore the frontiers that the latter have opened. Aside from
advancing abstract mathematics, the discovery of the ADS/CFT
conjecture \cite{Maldacena:2001uc} provides hope that results
within a (nonphysical) perturbative string theory may be
transformed to a mathematically dual (but physical)
nonperturbative theory, such as QCD. If true, this duality would
be a major breakthrough, and might by itself guarantee the
survival of string theory in some form, even if falsified by
experiment.

Studying the large number of theories in the landscape and how
they are related to each other has provided deep insights into how
a physical theory generally works. The string theory landscape may
be likened to a vast range of samples collected and studied in
detail for the purpose of understanding why theories of physics
behave the way they do and perhaps guide us into answering deep
questions about such things as symmetry and its origins.

So even if someone shows that the universe cannot be based on
string theory, I suspect that people will continue to work on it.
It might no longer be considered physics, nor will mathematicians
consider it to be pure mathematics. I can imagine that string
theory in that case may become its own new discipline; that is, a
mathematical science that is devoted to the study of the structure
of physical theory and the development of computational tools to
be used in the real world. The theory would be studied by
physicists and mathematicians who might no longer consider
themselves either. They will continue to derive beautiful
mathematical formulas and feed them to the mathematicians next
door. They also might, every once in a while, point out
interesting and important properties concerning the nature of a
physical theory which might guide the physicists exploring the
actual theory of everything over in the next building.

Whether or not string theory describes nature, there is no doubt
that we have stumbled upon an exceptionally huge and elegant
structure which might be very difficult to abandon. The formation
of a new science or discipline is something that happens
continually. For example, most statisticians do not consider
themselves mathematicians. In many academic institutions
departments of mathematics now call themselves ``mathematics and
statistics.'' Some have already detached into separate departments
of statistics. Perhaps the future holds a similar fate for the
unphysical as well as not-so-purely-mathematical new science of
string theory.

\end{document}